%% file: ARA_Veff_2025_ICRC.tex
\documentclass[a4paper,11pt]{article}

\usepackage{pos}
\usepackage{caption}
\usepackage{subcaption}
\usepackage{setspace}
\usepackage{titlesec}
\titlespacing{\section}{0pt}{*1.0}{*0.67}
\titlespacing{\subsection}{0pt}{*1.0}{*0.66}

\title{High-Fidelity Simulations of the Full Askaryan Radio Array and its Sensitivity to Ultra-High Energy Neutrinos}

\ShortTitle{High-Fidelity Simulations of ARA and its Sensitivity to UHE Neutrinos}

\author*[a,b]{Abigail Bishop} 
\author[c]{Alan Salcedo Gomez}
\author[a,b]{Marco Muzio}
\onbehalf{for the ARA Collaboration \\{\normalsize \normalfont(a complete list of authors can be found at the end of the proceedings)}\\}

\affiliation[a]{Wisconsin IceCube Particle Astrophysics Center,\\
Madison, WI 53703, USA}
\affiliation[b]{Dept. of Physics, University of Wisconsin-Madison,\\
Madison, WI 53706, USA}
\affiliation[c]{Dept. of Physics, Center for Cosmology and AstroParticle Physics, The Ohio State University,\\
Columbus, OH 43210, USA}

\emailAdd{abigail.bishop@wisc.edu}
\emailAdd{salcedogomez.1@osu.edu}
\emailAdd{muzio@wisc.edu}

\abstract{
  The Askaryan Radio Array (ARA) is a five-station, in-ice radio detector located at the South Pole searching for particle cascades from cosmogenic and astrophysical neutrinos with $\geq10^{17}$~eV of energy. 
  Cascades in this energy regime emit radio-wavelength Askaryan radiation that can be observed by one or more ARA stations. With the recent KM3Net observation of an approximately $220$~PeV neutrino, there is renewed, urgent interest in further unlocking the ultra-high energy neutrino sky. 
  We present updated calculations of ARA’s array-wide effective volume, sensitivity, and expected event rates for ultra-high energy neutrino-induced cascades. 
  Notably, results now account for the contributions of secondary particles from neutrino interactions (such as muon tracks) and multi-station detections within a detailed detector simulation framework. 
  Previous work has shown these secondary interactions and multi-station coincidences compose 25\% and 8\% of the detector’s effective area, respectively. 
  We intend to extend these results towards a novel analysis that estimates the degree to which secondary cascades and multi-station observations are detectable in a real neutrino search. 
  This will inform future UHE neutrino searches as it will characterize the feasibility of detecting such events.
}

\input{ICRCdetails.tex}

\begin{document}

\maketitle

\section{Introduction}\label{sec:intro}

  Ultra-high energy (UHE) neutrino astronomy ($E_{\nu}$ $\geq10^{17}$~eV) is pushing the boundaries of high energy astrophysics, especially given recent developments in instrumentation, analysis, and the recent KM3NeT discovery of an approximately $110-790$~PeV neutrino~\cite{km3net}.
  Ice-focused UHE neutrino astronomy is an extremely active area of research with dedicated experiments in both the Artic and Antarctic.
  For example, the Askaryan Radio Array (ARA) is a 5-station in-ice array at the South Pole conducting an array-wide and all-livetime search for neutrinos through 11 years of data \cite{ara_5sa_icrc2025}. 
  The Radio Neutrino Observatory in Greenland is constructing a 35-station in-ice array and is performing a neutrino search with data taken by the 7 of the 8 stations currently installed \cite{rnog_7sa_icrc2025}. 
  The Payload for Ultrahigh Energy Observations is a next generation balloon-based experiment planned to launch over Antarctica this December and January to collect data for a neutrino search  \cite{peuo_icrc2025}. 
  This balloon's precursor, the ANtarctic Impulsive Transient Antenna (ANITA), flew 4 times and identified several events that appear to be inconsistent with expected background behavior and continue to interest the community and motivate searches from experiments such as the Pierre Auger Observatory \cite{anita_2016, anita_2018, anita_auger}.
  Considering the community intrigue in UHE neutrino astronomy and a lack of published neutrino observations over $10^{18}$~eV, despite multiple ice-based searches with ARA, ANITA, and other detectors, having increasingly thorough simulations is essential to accurately determine experimental sensitivities to the UHE neutrino diffuse flux.
  In this work, we use ARA's simulation framework, AraSim, to perform an updated and increasingly thorough simulation that will be used in ARA's array-wide and all-livetime neutrino search~\cite{ara_a23_4yr}. 
  Previous sensitivities are likely under-estimated because all possible event signatures, such as muon and tau tracks, were not simulated. 
  Recent results using NuLeptonSim and Python for Radio Experiments showed that, for ARA, events observed via outgoing particles from neutrino cascades, namely muon tracks and tau decays, comprise 25\% of expected observations and multi-station events make up 8\%~\cite{ara_secondaries_icrc2023}.
  Simulations are also passed through the analysis pipeline allowing for both a trigger-level and an analysis-level sensitivity study.

    \begin{figure}
    
      \hspace{0.015\textwidth}
      \begin{subfigure}[t]{0.45\textwidth}
        \centering
        \includegraphics[width=2.5in]{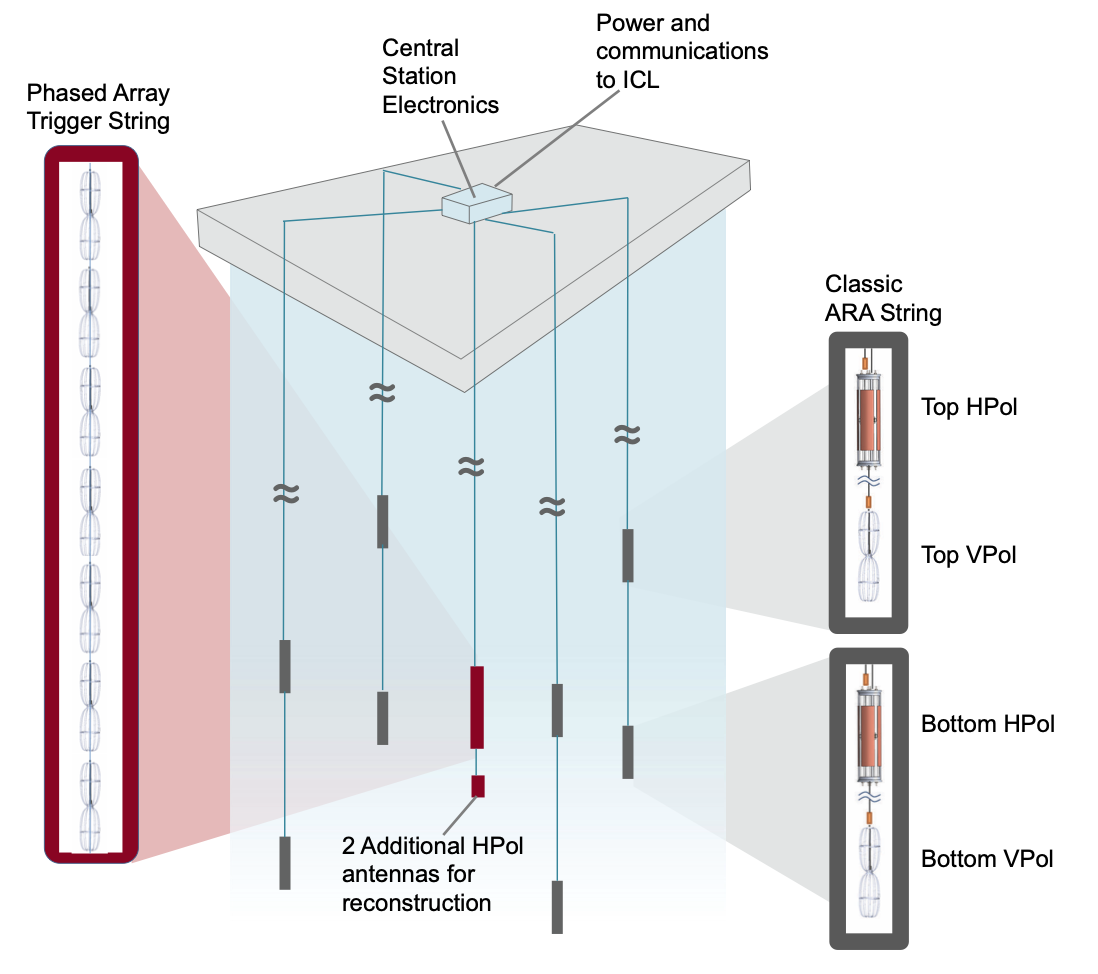}
        \caption{}
        \label{fig:ara_station}
      \end{subfigure}
      \hspace{0.035\textwidth}
      \begin{subfigure}[t]{0.45\textwidth}
        \centering
        \includegraphics[width=2.5in]{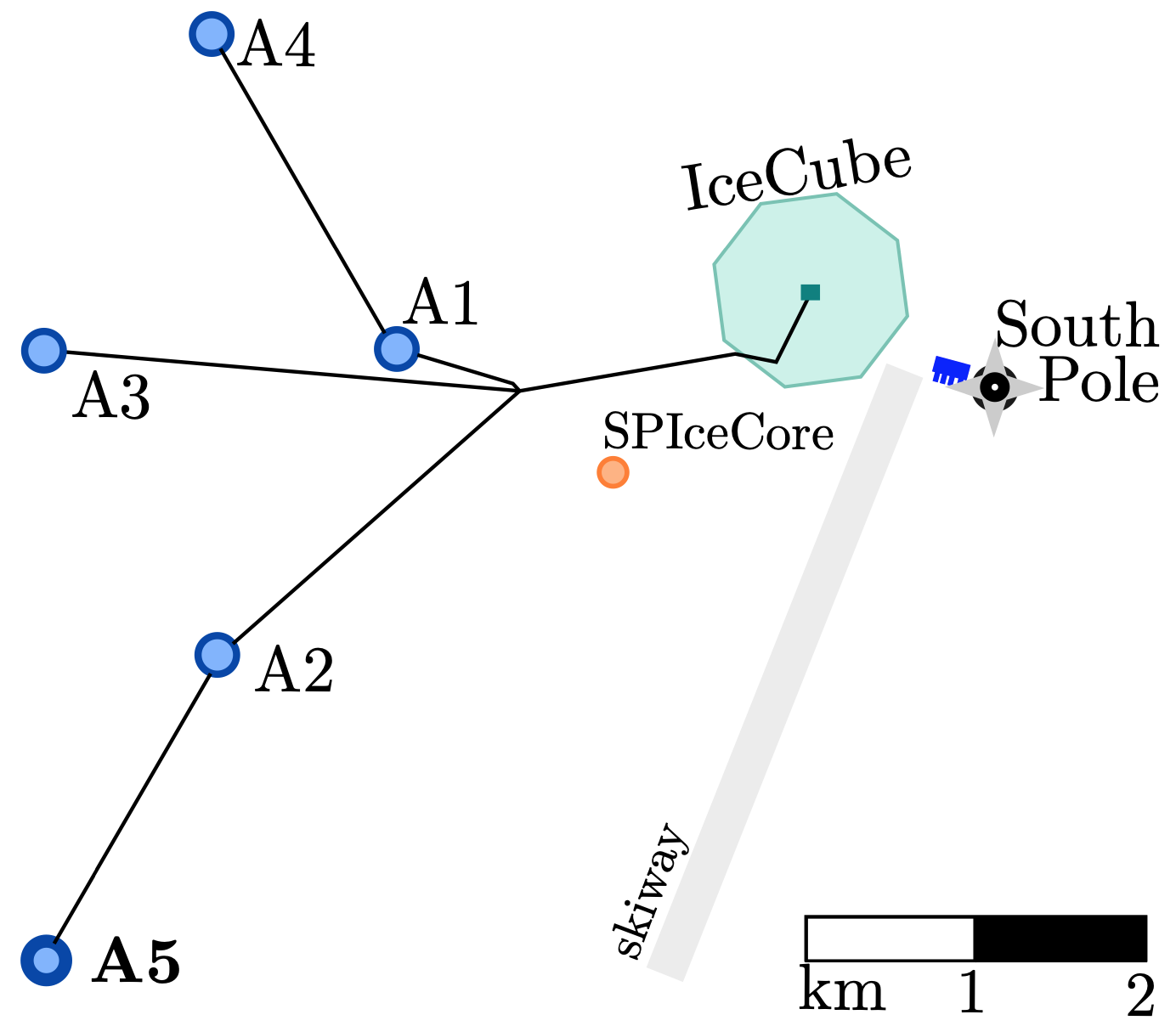}
        \caption{}
        \label{fig:ara_array}
      \end{subfigure}
      
      \caption{
        Left is the station design of an Askaryan Radio Array station with four outer, standard strings and one Phased Array string located in the center. A5 has this layout while the other four stations only have the 4 outer strings. Right is the layout of all stations in the Askaryan Radio Array alongside local South Pole landmarks.
      }
      \label{fig:station}
      
    \end{figure}

    \begin{table}[!ht]
      \centering
      \begin{tabular}{| c | c | c | c | c |}
        \hline
        Station & Installation & Deployment & Outer String& Working \\
        Name & Year & Depth [m] & Spacing [m] & Strings \\ 
        \hline
        A1 & 2012 & 77 & 14 & 4 \\
        A2 & 2013 & 191 & 14 & 4 \\
        A3 & 2013 & 193 & 14 & 4 \\
        A4 & 2018 & 193 & 28 & 3 \\
        A5 and PA & 2018 & 186 & 33 & 5 \\
        \hline
      \end{tabular}
      \caption{For each station, the year of installation, average maximum depth of all strings in the station, the average baseline X-Y string spacing, and the number of data-taking strings.}
      \label{tab:ara_stations}
    \end{table}

  ARA's five stations are spaced 2 kilometers apart atop the roughly 3 kilometer deep glacier at the South Pole. 
  Each station has 8 pairs of antennas sensitive to vertically polarized  (Vpol) and horizontally polarized (Hpol) electric fields. 
  The pairs of antennas are installed as a cubic lattice in the ice, as shown on the four outer strings in Figure~\ref{fig:ara_station}. 
  %This was done by deploying 4 strings in the Antarctic glacier arranged in a square with a predetermined baseline length and two antenna pairs separated by the same baseline length. 
  A summary of the five stations is presented in Table~\ref{tab:ara_stations} as some key features affect the results presented in this proceeding.
  A5 has an additional, central string, called the Phased Array (PA), instrumented with 7 closely packed Vpol antennas above 2 Hpol antennas deployed to a depth around 190\,m.
  The PA uses beamforming techniques to trigger on events with significantly lower signal-to-noise ratios than the traditional ARA stations but can only reconstruct the observed zenith angle of the interaction vertex since all antennas are vertically stacked. 
  To capitalize on the increased sensitivity provided by its trigger, the PA forces readout of the A5 detector to create a hybrid detection channel with full vertex reconstruction capabilities due to the azimuthal sensitivity provided by the A5 antennas. 
  When counting ARA stations, the collocated A5 and PA detectors are considered to makeup one station, since in many scenarios both the A5 and PA detectors are expected to observe the same events. 
  In this proceeding, we will at times consider the performance of the PA separately from A5 because the A5 DAQ records all PA-triggered events, but PA does not record events triggered by A5. 

\section{Simulation}\label{sec:sim}

  \begin{figure}
    \centering
    \includegraphics[width=2.25in]{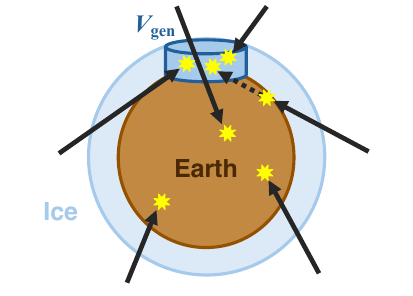}
    \caption{Drawing of neutrino trajectories (solid lines) and a trajectory of an outgoing particle (dashed line) that all pass through the cylindrical generation volume (blue, $V_{\text{gen}}$) and vertices (yellow stars) where each particle interacted with the Earth (brown) or the shell of ice (light blue). This diagram is not to scale. In this example, 6 events are thrown by NuLeptonSim but only 3 cascades fall in $V_{\text{gen}}$ and would be simulated with AraSim.}
    \label{fig:NLSEarth}
  \end{figure}
  
  The simulation has two main steps. 
  The first main step is the physics simulation of the neutrino interaction. 
  % Reuse the ICRC 2023 diagram of event topologies and explain some of the key topologies
  We run the lepton propagation framework NuLeptonSim to generate initial neutrino interactions, muon and tau stochastic energy losses (collectively referred to as tracks), and all outgoing tau and secondary neutrino interactions \cite{nls}. 
  The energy and inelasticity of each induced shower are saved along with the shower's direction and a unique primary particle identifier that allows us to connect all related cascades in the detector simulation.
  Earth absorption is modeled according to the Preliminary Earth Reference Model with a 3 kilometer thick shell of ice modeled with the equation below \cite{prem, ara_pa}. 
  \begin{equation}
    n(z) = 1.780 - 0.454 e^{(-0.0202)\times \frac{z}{1~\text{m}}}
    \label{eq:ice}
  \end{equation}
  A generation volume encapsulating ARA is defined as a 15\,km radius and 3\,km depth cylinder centered on A2. 
  A random location is chosen in this generation volume then a direction of travel is chosen by uniformly sampling azimuth from 0 to $2\pi$ and zenith in $\cos(\theta)$ space from -1 to 1.
  %A chord is drawn through the Earth to the point on the surface where a neutrino traveling in this direction would've first entered Earth. 
  An exaggerated diagram of the generation volume, the Earth model, and examples of a few events are shown in Figure~\ref{fig:NLSEarth}. 
  Along each trajectory, the code then forward propagates neutrinos and samples an interaction point based on the neutrino's energy, flavor, and cross section.
  An equal number of neutrinos are simulated per flavor and energy. 
  The 8 energies simulated for this work range from $10^{17.5}$ eV to $10^{21}$ eV spaced in half energy decades. 
  Any muons, taus, and neutrinos leaving the initial cascade with energy greater than $10^{16}$ eV are allowed to travel and interact further along the trajectory. 
  The muons and taus release energy in the form of stochastic losses until they decay, leave the Earth, or their energy falls below $10^{16}$~eV.
  Cascades from initial neutrinos, tracks, tau decays, and outgoing neutrino interactions that fall within the generation volume are then saved to an event list. 

  The second main step is simulating the detector response to all cascades in the event list with AraSim. 
  For each cascade, its radio signal is propagated through the ice to an antenna, where it is convolved with the antenna's response function and the detector's electronics gain response, resulting in a voltage waveform.
  This ARA-specific simulation has undergone multiple major developments to prepare for this simulation setup. 
  Developments include updated data-driven noise and electronic gain models along with an improved PA trigger simulation.
  A new pipeline was developed to more accurately simulate the observation of multiple cascades (from a muon track, for example) and  multiple triggers generated by primary and secondary cascades while properly accounting for detector deadtime after triggering.
  If the station triggers and there is still waveform left to analyze after the initial readout and subsequent station deadtime, the remaining waveform will be analyzed and the station will be allowed to trigger a second time on different cascades in the same event. 
  % talk about how often this would happen given the back of the envelope we did

\section{Calculating Effective Volume}

  There are a couple subtleties to this work's simulation setup that require extra care when calculating the effective volume. 
  Firstly, typical neutrino simulations force interactions within the generation volume whereas this work allows particles to forward propagate, changing how interaction and survival probabilities will be used.
  When neutrinos are forced to interact, interaction weights estimate the probability that an event interacts at its chosen vertex depending on its total chord length and distance traveled through the generation volume.
  The survival weight estimates the probability that neutrinos arriving from a particular zenith direction have not interacted with any nucleons by the time they enter the generation volume.
  Secondly, the anisotropic sampling of neutrino directions needs to be accounted for. 
  When neutrino events are chosen, their direction of travel is chosen randomly in $4\pi$ space and not randomly with respect to their entrance and exit points in the generation volume or Earth. 
  This is typically encapsulated in the interaction probability and will have to be employed here. 
  This probability calculated is that of the initially generated neutrino but is also applied to all station triggers associated with the same initial neutrino.
  
  The basic effective volume calculation follows this form
  \begin{equation}
    [V \Omega]_{\text{eff}}(E) = \int_{\Omega} d\Omega \int_V d^3r P_{\text{surv}}(E, \Omega, r) ~ P_{\text{trig}}(E, \Omega, r)
    \label{eq:veff_def_int}
  \end{equation}
  where $\Omega$ is the solid angle, $V$ is the generation volume in which events are passed through the detector simulation, $P_{\text{surv}}$ is a particle's survival probability, and $P_{\text{trig}}$ is a particle's triggering probability. 
  This effective volume can be connected to the effective area by multiplying by the particle's interaction probability, $P_{\text{int}}=1/L_{\text{int}}$.
  The solid angle integration usually results in a scaling factor of $4\pi$, the volume integration results in the volume of the generation volume, and the probability of triggering is evaluated by a simulation framework and can be replaced by $\delta_{\text{trig}}=0,1$ to obtain
  \begin{equation}
    [V \Omega]_{\text{eff}} \simeq \left( \int_{\Omega_{\text{gen}}} d\Omega \int_{
    V_{\text{gen}}} dV \right) \frac{1}{N} \sum^N_{i=1} P_{\text{surv}, i} ~ \delta_{\text{trig}, i} ~\text{.}
  \end{equation}
  Since interaction vertices in the generation volume are conditioned on surviving to and interacting in this volume, their distribution is not uniform and isotropic in the volume. 
  This amounts to an importance sampling of the volume with points drawn proportional to  
  \begin{equation}
    P_{\text{surv}}(E, \Omega)P_{\text{int}}~.
    \label{eq:rejectionsampling}
  \end{equation}
  Therefore, the effective volume equation becomes 
  \begin{equation}
    [V \Omega]_{\text{eff}} \simeq \left( \int_{\Omega_{\text{gen}}} d\Omega \int_{
    V_{\text{gen}}} dV P_{\text{surv}}(E, \Omega)P_{\text{int}} \right) \frac{1}{N} \sum^N_{i=1} \frac{P_{\text{surv}, i} ~  \delta_{\text{trig}, i}}{P_{\text{surv}, i}P_{\text{int}} } ~ \text{.}
    \label{eq:veff3}
  \end{equation}
  We calculate the product of the survival and interaction probabilities numerically by using the event generator to throw events and by tracking how many particles survive to reach the 15 km generation volume. 
  The ratio of the number of neutrinos with a direction $\cos\theta$ with an initial and/or secondary cascade interacting within the generation volume compared to the number of total thrown neutrinos is then used as an approximation for the rejection sampling in Equation~\ref{eq:rejectionsampling}, $w(\cos\theta)$.
  \begin{equation}
    P_{\text{surv}}(E, \Omega)P_{\text{int}}  \simeq w(\cos\theta) \equiv \frac{N_{\text{gen}}(\cos\theta)}{N(\cos\theta)}
  \end{equation}
  We approximate the interaction probability similarly by generating events and by calculating how many initial neutrinos interact within the generation volume compared to how many pass through without interacting.
  By substituting this into Equation~\ref{eq:veff3}, integrating over the generation volume, and moving the remaining $P_{\text{surv}, i}$ term into the denominator of the sum we get
  \begin{equation}
    [V \Omega]_{\text{eff}} \simeq \left( \int_{\Omega_{\text{gen}}} w(\cos\theta) ~ d\Omega \int_{V_{\text{gen}}} dV \right) \frac{1}{N} \sum^N_{i=1} \frac{ \delta_{\text{trig}, i}}{w(\cos\theta) ~/~ P_{\text{surv}, i}(\cos\theta)}
    \label{eq:veff}
  \end{equation}
  where $P_{\text{surv}}( \cos\theta)$ is also approximated as the number of neutrino events where the neutrino or one of its secondary particles survives Earth absorption enough to pass through the generation volume. 
  We then use a Monte Carlo simulation to calculate $w(\cos\theta)$ and $w(\cos\theta)/P_{\text{surv}, i}(\cos\theta)$ for various directional zenith bins to use when calculating the final effective volume in Equation~\ref{eq:veff}.

  ARA has 39 different livetime configurations driven by which individual detectors are turned on at any point and what configuration each detector is in. 
  To better characterize the performance of the array, we calculate the effective volume for each livetime configuration and weight the result by the duration of that configuration: 

  \begin{equation}
    [V\Omega]_{\text{eff}}(E) = \frac{1}{T_{\text{tot}}} \sum_c T_c~[V\Omega]_{\text{eff}, c}(E) ~ \text{.}
  \end{equation}

  \noindent This produces an exposure-like effective volume that accurately the large effective volume from having 5 detectors online with less optimal livetime configurations.

  % Need to add a section about livetime averaging
  % Mention that this turns it into an exposure almost

\section{Results}

    \begin{figure}[!ht]
      \centering
      \includegraphics[width=2.5in]{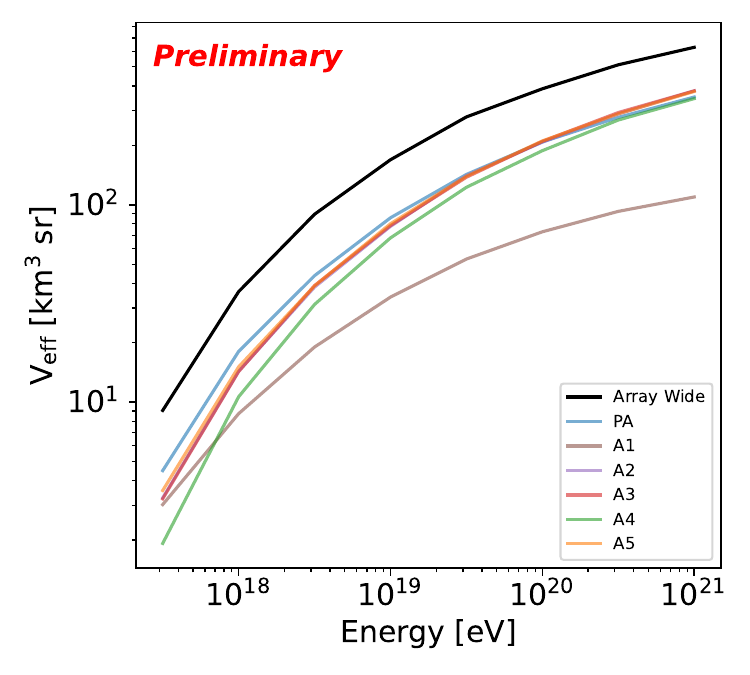}
      \caption{In black, the array-wide and livetime-weighted effective volume of ARA. Colored lines show the effective volume of each station (not scaled for livetime), separating the collocated A5 and PA to show the power of a single-string phased array detector.}
      \label{fig:veffs}
    \end{figure}

    Figure~\ref{fig:veffs} shows the array-wide livetime-weighted effective volume of the full ARA in black. 
    Each colored line shows the effective volumes for each subdetector of ARA, demonstrating A5 and the PA separately. 
    This plot shows that the array-wide effective volume is 2-3 times larger than the average single-station effective volume. 
    We also consider the distinctions between stations summarized in Table~\ref{tab:ara_stations} compared to Figure~\ref{fig:veffs}. 
    A1, deployed half as deep as all other stations, has less sensitivity than the other stations at high energies but increasing sensitivity at low energies. 
    A4, which only has 3 strings, is less effective at low energies compared to the single-string PA and all other 4-string stations.
    Lastly, the PA has greater sensitivity to neutrinos at lower energies but is comparable to the 4-string stations at highest energies. 

    \begin{figure}
    
      \hspace{0.015\textwidth}
      \begin{subfigure}[t]{0.45\textwidth}
        \centering
        \includegraphics[width=2.5in]{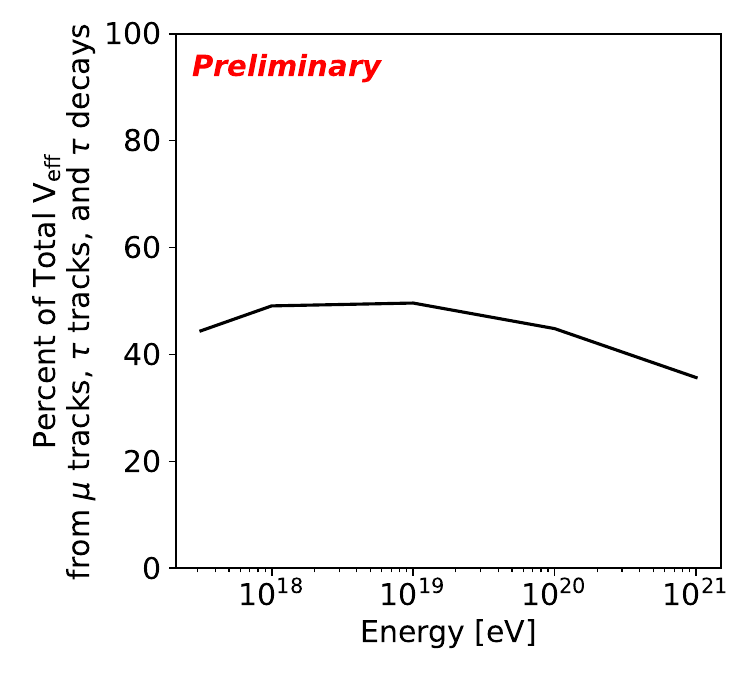}
        \label{fig:wwosecondaries}
      \end{subfigure}
      \hspace{0.035\textwidth}
      \begin{subfigure}[t]{0.45\textwidth}
        \centering
        \includegraphics[width=2.5in]{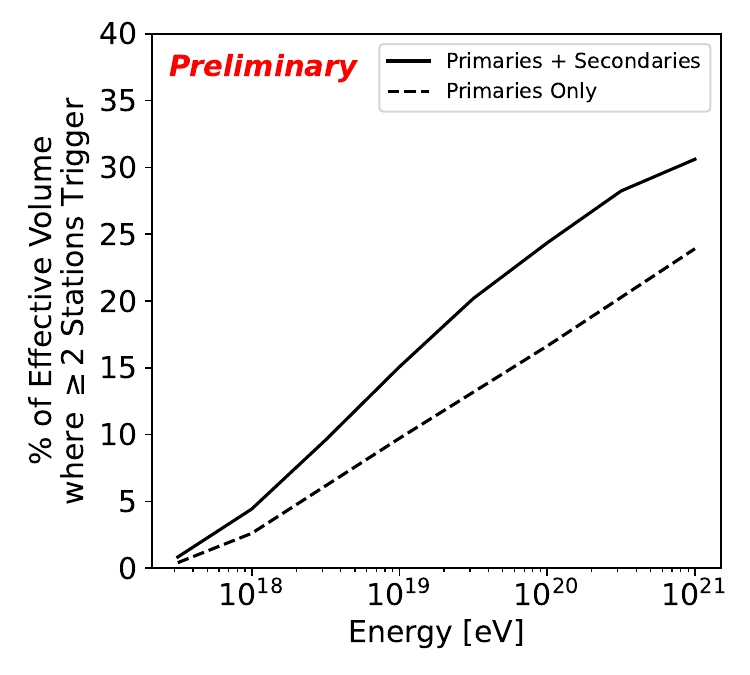}
        \label{fig:multistation}
      \end{subfigure}
      
      \caption{Left: Percent of ARA's array-wide effective volume comprised of observations of secondary particle cascades, including muon tracks, tau tracks, and tau decays. Right: Percent of ARA's effective volume from events where two or more stations trigger on one or more cascades connected to the same primary neutrino. For example, separate cascades from the same muon track triggering separate detectors or a tau neutrino and a subsequent tau decay would contribute to the multi-station rate. }
      \label{fig:veff_breakdown}
    \end{figure}

  The left plot of Figure~\ref{fig:veff_breakdown} shows the percent of the total effective volume comprised of events where a track, tau decay, or outgoing neutrino triggered one of the ARA stations but not the initial cascade from a primary neutrino. 
  This should signify the percent of ARA's total effective volume that was not previously simulated before the addition of the framework outlined in Section~\ref{sec:sim}.
  This implies that almost half of triggering events in ARA are expected to be from secondary particle cascades.
  The right plot of Figure~\ref{fig:veff_breakdown} shows the percent of the total effective volume made up of events that trigger two or more stations. 
  This rate increases with energy and when secondary particles are added to the simulation. 
  Previous results concluded that roughly 5\% of $10^{18}$~eV primary neutrino interactions triggered both A2 and A3, but we calculate this rate to be 2.6\% for the full array when considering only primary neutrino interactions \cite{ara_a23_10mo}.
  When adding in the secondary interactions, this number increases to 4.4\% at $10^{18}$~eV. 

  The predicted sensitivity of ARA is displayed in Figure~\ref{fig:sensitivity} along with the sensitivities of other experiments and a selection of theoretical flux models.
  This sensitivity is calculated using the array-wide effective volume shown in Figure~\ref{fig:veffs}, the livetime of the full array (10.6 years), and the analysis efficiencies from the neutrino search covering 4 years of A2 and A3 data \cite{ara_a23_4yr}. 
  These analysis efficiencies are expected to be conservative compared to the analysis efficiencies of the upcoming array-wide analysis which uses state of the art data cleaning and signal-background discrimination algorithms \cite{ara_5sa_icrc2025}. 
  These results indicate that ARA is sensitive to the flux inferred by the KM3NeT event, a cosmogenic maximum proton model \cite{muzio_flux}, and that our single event sensitivity can probe a star formation rate flux with $E_{\text{max}}=10^{21.5}$ eV \cite{koteraetal}.
  At the completion of ARA's array-wide analysis, ARA will be the most sensitive detector to neutrinos with energies greater than at least $10^{10}$ GeV. 

  \begin{center}
    \begin{minipage}{0.45\textwidth}
      % \vspace{0.2in}
      \centering
      \includegraphics[width=\textwidth]{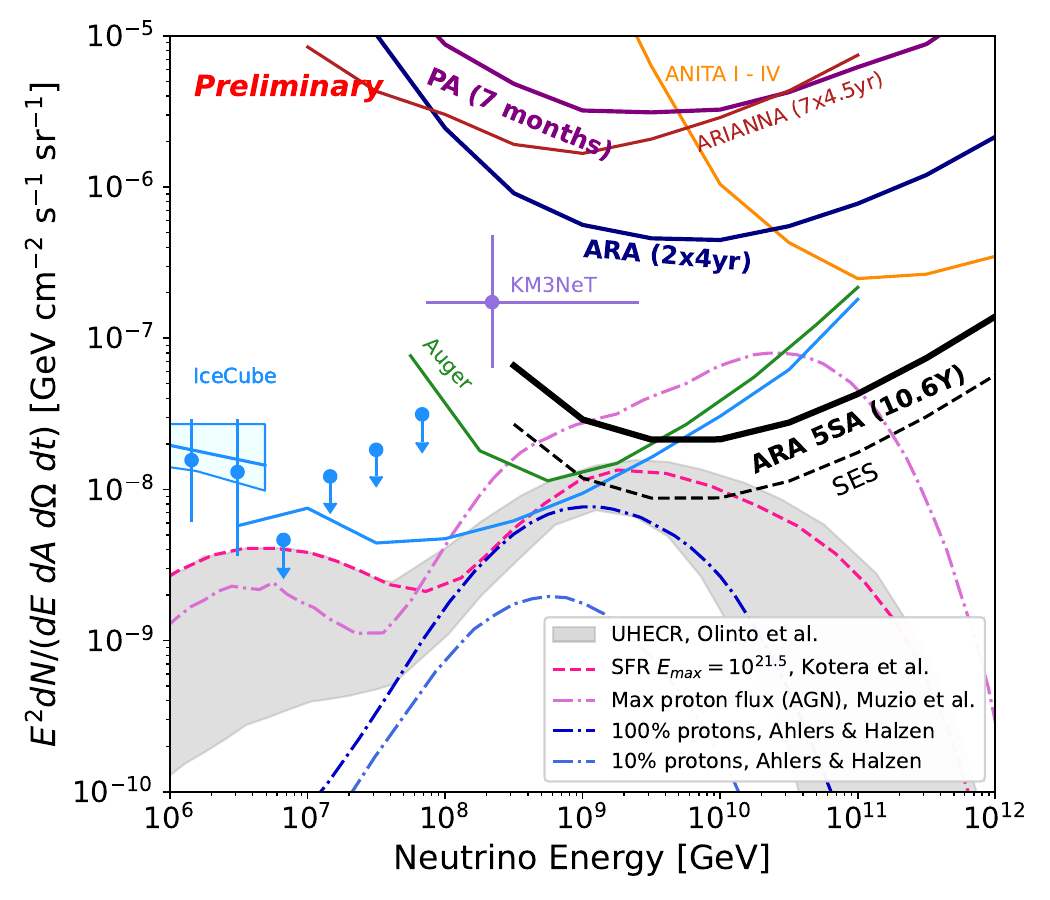}
    \end{minipage}
    \begin{minipage}{0.35\textwidth}
      \footnotesize
      \begin{center}
        Number of events in 25 station-years:
        \hspace{24pt}
        \begin{tabular}{ | p{1.55in} | c | }
          \hline
          \textbf{Flux} & \textbf{$N_{\text{events}}$} \\ \hline 
          IceCube EHE 2025 \cite{icecube_ehe_2025} + ANITA I-IV \cite{anita_iv_flux} & 39.2 \\ \hline
          Muzio et al \cite{muzio_flux} & 28 \\ \hline
          Kotera High E \cite{koteraetal} & 5.3 \\ 
          \hline
        \end{tabular}
      % \captionof{table}{yk}
      \end{center}
    \end{minipage}
    \captionof{figure}{Left: The predicted sensitivity of ARA from 2012 through 2023. The solid, black line shows the expected sensitivity with a Feldman-Cousins coefficient of 2.44 representing the sensitivity of a data sample with 0 background events and the dashed black line shows the sensitivity with a Feldman-Cousins coefficient of 1, representing the single event sensitivity. Right: Expected event rates for the expected ARA sensitivity computed with an assortment of flux models. }
    \label{fig:sensitivity}
  \end{center}

\section{Conclusion}\label{sec:conclusion}

  The Askaryan Radio Array is a large in-ice neutrino detector made up of 5 stations at the South Pole that has been taking data since 2012. This work showcases the predicted effective volume, sensitivity, and event rates using a simulation framework updated to include all tracks, tau decays, and outgoing neutrinos resulting from initial neutrino interactions.
  An analysis of all the data taken up to December 2023 is underway and will demonstrate, given the results in this proceeding, that ARA is the most sensitive neutrino detector for neutrinos with energy greater than a few EeV. 
  The sensitivity of ARA will probe theoretical models of the ultra-high energy neutrino flux and the flux inferred by KM3NeT's observation. 
  The upcoming neutrino search in 10.6 years of data array-wide (or 27 station-years) will be the most powerful neutrino search until at least 2030 as other detectors continue to be built and accumulate data. 

% Bibtex references:
\begingroup
\setstretch{0.5}
\setlength{\bibsep}{1.0pt}
\bibliographystyle{ICRC}
\bibliography{references}
\endgroup

% Alternatively, you can include references by hand:
%\begin{thebibliography}{99}
%\bibitem{...}
%
%\end{thebibliography}

\clearpage

%The following list of authors, affiliations and funding agencies should be updated at the day of submission. 
\input{ara_icrc_authors}

\input{ara_icrc25_acknowledgements}

\end{document}

%% file: ICRCdetails.tex
\ConferenceLogo{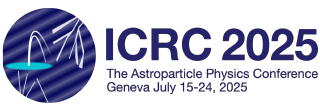}

\FullConference{39th International Cosmic Ray Conference (ICRC2025)\\
 15–24 July 2025\\
Geneva, Switzerland\\}

%% file: ara_icrc_authors.tex
% ICRC list for ARA Collaboration
\section*{Full Author List: ARA Collaboration (June 30, 2025)}

\noindent
N.~Alden\textsuperscript{1}, 
S.~Ali\textsuperscript{2}, 
P.~Allison\textsuperscript{3}, 
S.~Archambault\textsuperscript{4}, 
J.J.~Beatty\textsuperscript{3}, 
D.Z.~Besson\textsuperscript{2}, 
A.~Bishop\textsuperscript{5}, 
P.~Chen\textsuperscript{6}, 
Y.C.~Chen\textsuperscript{6}, 
Y.-C.~Chen\textsuperscript{6}, 
S.~Chiche\textsuperscript{7}, 
B.A.~Clark\textsuperscript{8}, 
A.~Connolly\textsuperscript{3}, 
K.~Couberly\textsuperscript{2}, 
L.~Cremonesi\textsuperscript{9}, 
A.~Cummings\textsuperscript{10,11,12}, 
P.~Dasgupta\textsuperscript{3}, 
R.~Debolt\textsuperscript{3}, 
S.~de~Kockere\textsuperscript{13}, 
K.D.~de~Vries\textsuperscript{13}, 
C.~Deaconu\textsuperscript{1}, 
M.A.~DuVernois\textsuperscript{5}, 
J.~Flaherty\textsuperscript{3}, 
E.~Friedman\textsuperscript{8}, 
R.~Gaior\textsuperscript{4}, 
P.~Giri\textsuperscript{14}, 
J.~Hanson\textsuperscript{15}, 
N.~Harty\textsuperscript{16}, 
K.D.~Hoffman\textsuperscript{8}, 
M.-H.~Huang\textsuperscript{6,17}, 
K.~Hughes\textsuperscript{3}, 
A.~Ishihara\textsuperscript{4}, 
A.~Karle\textsuperscript{5}, 
J.L.~Kelley\textsuperscript{5}, 
K.-C.~Kim\textsuperscript{8}, 
M.-C.~Kim\textsuperscript{4}, 
I.~Kravchenko\textsuperscript{14}, 
R.~Krebs\textsuperscript{10,11}, 
C.Y.~Kuo\textsuperscript{6}, 
K.~Kurusu\textsuperscript{4}, 
U.A.~Latif\textsuperscript{13}, 
C.H.~Liu\textsuperscript{14}, 
T.C.~Liu\textsuperscript{6,18}, 
W.~Luszczak\textsuperscript{3}, 
A.~Machtay\textsuperscript{3}, 
K.~Mase\textsuperscript{4}, 
M.S.~Muzio\textsuperscript{5,10,11,12}, 
J.~Nam\textsuperscript{6}, 
R.J.~Nichol\textsuperscript{9}, 
A.~Novikov\textsuperscript{16}, 
A.~Nozdrina\textsuperscript{3}, 
E.~Oberla\textsuperscript{1}, 
C.W.~Pai\textsuperscript{6}, 
Y.~Pan\textsuperscript{16}, 
C.~Pfendner\textsuperscript{19}, 
N.~Punsuebsay\textsuperscript{16}, 
J.~Roth\textsuperscript{16}, 
A.~Salcedo-Gomez\textsuperscript{3}, 
D.~Seckel\textsuperscript{16}, 
M.F.H.~Seikh\textsuperscript{2}, 
Y.-S.~Shiao\textsuperscript{6,20}, 
S.C.~Su\textsuperscript{6}, 
S.~Toscano\textsuperscript{7}, 
J.~Torres\textsuperscript{3}, 
J.~Touart\textsuperscript{8}, 
N.~van~Eijndhoven\textsuperscript{13}, 
A.~Vieregg\textsuperscript{1}, 
M.~Vilarino~Fostier\textsuperscript{7}, 
M.-Z.~Wang\textsuperscript{6}, 
S.-H.~Wang\textsuperscript{6}, 
P.~Windischhofer\textsuperscript{1}, 
S.A.~Wissel\textsuperscript{10,11,12}, 
C.~Xie\textsuperscript{9}, 
S.~Yoshida\textsuperscript{4}, 
R.~Young\textsuperscript{2}
\\
\\
\textsuperscript{1} Dept. of Physics, Enrico Fermi Institute, Kavli Institute for Cosmological Physics, University of Chicago, Chicago, IL 60637\\
\textsuperscript{2} Dept. of Physics and Astronomy, University of Kansas, Lawrence, KS 66045\\
\textsuperscript{3} Dept. of Physics, Center for Cosmology and AstroParticle Physics, The Ohio State University, Columbus, OH 43210\\
\textsuperscript{4} Dept. of Physics, Chiba University, Chiba, Japan\\
\textsuperscript{5} Dept. of Physics, University of Wisconsin-Madison, Madison,  WI 53706\\
\textsuperscript{6} Dept. of Physics, Grad. Inst. of Astrophys., Leung Center for Cosmology and Particle Astrophysics, National Taiwan University, Taipei, Taiwan\\
\textsuperscript{7} Universite Libre de Bruxelles, Science Faculty CP230, B-1050 Brussels, Belgium\\
\textsuperscript{8} Dept. of Physics, University of Maryland, College Park, MD 20742\\
\textsuperscript{9} Dept. of Physics and Astronomy, University College London, London, United Kingdom\\
\textsuperscript{10} Center for Multi-Messenger Astrophysics, Institute for Gravitation and the Cosmos, Pennsylvania State University, University Park, PA 16802\\
\textsuperscript{11} Dept. of Physics, Pennsylvania State University, University Park, PA 16802\\
\textsuperscript{12} Dept. of Astronomy and Astrophysics, Pennsylvania State University, University Park, PA 16802\\
\textsuperscript{13} Vrije Universiteit Brussel, Brussels, Belgium\\
\textsuperscript{14} Dept. of Physics and Astronomy, University of Nebraska, Lincoln, Nebraska 68588\\
\textsuperscript{15} Dept. Physics and Astronomy, Whittier College, Whittier, CA 90602\\
\textsuperscript{16} Dept. of Physics, University of Delaware, Newark, DE 19716\\
\textsuperscript{17} Dept. of Energy Engineering, National United University, Miaoli, Taiwan\\
\textsuperscript{18} Dept. of Applied Physics, National Pingtung University, Pingtung City, Pingtung County 900393, Taiwan\\
\textsuperscript{19} Dept. of Physics and Astronomy, Denison University, Granville, Ohio 43023\\
\textsuperscript{20} National Nano Device Laboratories, Hsinchu 300, Taiwan\\

%% file: ara_icrc25_acknowledgements.tex
\section*{Acknowledgements}

\noindent
The ARA Collaboration is grateful to support from the National Science Foundation through Award 2013134.
The ARA Collaboration
designed, constructed, and now operates the ARA detectors. We would like to thank IceCube, and specifically the winterovers for the support in operating the
detector. Data processing and calibration, Monte Carlo
simulations of the detector and of theoretical models
and data analyses were performed by a large number
of collaboration members, who also discussed and approved the scientific results presented here. We are
thankful to Antarctic Support Contractor staff, a Leidos unit 
for field support and enabling our work on the harshest continent. We thank the National Science Foundation (NSF) Office of Polar Programs and
Physics Division for funding support. We further thank
the Taiwan National Science Councils Vanguard Program NSC 92-2628-M-002-09 and the Belgian F.R.S.-
FNRS Grant 4.4508.01 and FWO. 
K. Hughes thanks the NSF for
support through the Graduate Research Fellowship Program Award DGE-1746045. A. Connolly thanks the NSF for
Award 1806923 and 2209588, and also acknowledges the Ohio Supercomputer Center. S. A. Wissel thanks the NSF for support through CAREER Award 2033500.
A. Vieregg thanks the Sloan Foundation and the Research Corporation for Science Advancement, the Research Computing Center and the Kavli Institute for Cosmological Physics at the University of Chicago for the resources they provided. R. Nichol thanks the Leverhulme
Trust for their support. K.D. de Vries is supported by
European Research Council under the European Unions
Horizon research and innovation program (grant agreement 763 No 805486). D. Besson, I. Kravchenko, and D. Seckel thank the NSF for support through the IceCube EPSCoR Initiative (Award ID 2019597). M.S. Muzio thanks the NSF for support through the MPS-Ascend Postdoctoral Fellowship under Award 2138121. A. Bishop thanks the Belgian American Education Foundation for their Graduate Fellowship support.